# On the Hardness of Boron (III) Oxide

V. A. Mukhanov, O. O. Kurakevich, and V. L. Solozhenko

*LPMTM-CNRS, Université Paris Nord, Villetaneuse, France*

At present two crystalline modifications, $\alpha$-$B_2O_3$ (a low-pressure phase) and $\beta$-$B_2O_3$ (a high-pressure phase), and g-$B_2O_3$ amorphous (glass-like) phase of boron (III) oxide are known [1]. The local structure of g-$B_2O_3$ is most close to the structure of the $\alpha$-modification [2], which crystallizes in the $P$31 space group with the $a$ = 4.3358 Å and $c$ = 8.3397 Å lattice parameters [3] (Fig. 1a). The denser $\beta$-modification (the $Cmc$2 space group, $a$ = 4. 613 Å, $b$ = 7.803 Å, $c$ = 4.129 Å [4]) (Fig. 1b) forms by crystallization from a melt at pressures above 7 GPa. There is no data in the literature on the boron oxide hardness, however, a high ($K$ = 180 GPa [5]) bulk modulus of $\beta$-$B_2O_3$ allows us to anticipate that this phase is of high hardness. In the present letter we report the measurements of hardness of g- and $\beta$-$B_2O_3$ phases.

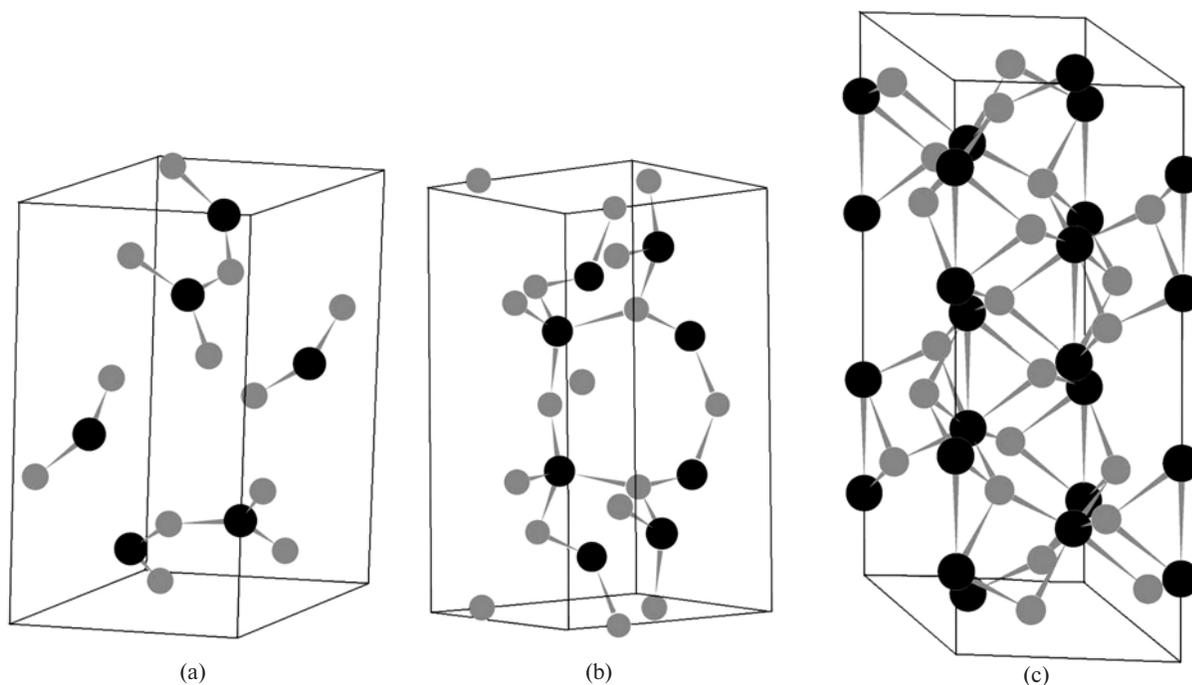

(a)          (b)          (c)

**Fig. 1**. Crystalline structures of $\alpha$-$B_2O_3$ (a), $\beta$-$B_2O_3$ (b) and $\gamma$-$B_2O_3$ hypothetical dense phase with a corundum structure (c); black and grey balls indicate boron and oxygen atoms, respectively.

Glass-like $B_2O_3$ was produced by decomposing metaboric acid $HBO_2$ at 940 K and subsequent remelting in order to remove air bubbles. The diffraction pattern of the as-synthesized sample exhibits two characteristic wide halos with $d_{hkl} \approx 3.5$ and 2.0 Å (Fig. 2a). The $B_2O_3$ high-pressure phase was synthesized from remelted g-$B_2O_3$ in a toroid-type apparatus at 7.2 GPa and 1020–1400 K. The X-ray diffraction studies of the samples (Fig. 2b) show that the samples are highly crystalline $\beta$-$B_2O_3$ [4] without impurities of foreign phases.

Vickers hardness was measured using a Duramin-20 (Struers) microhardness tester at indentation loads from 0.5 to 20 N and a holding time of 20 s. At each load at least four indents were placed at about 200 μm intervals.

Our findings show (Fig. 3) that the hardness of glass-like $B_2O_3$ is of about 1.5 GPa, while the hardness of the high-pressure phase is higher by a factor of 10 (16 ± 5 GPa) and comparable with the hardness (16 GPa) of the WC–10% Co hard alloy [6]. Since for the $B_2O_3$ composition, the structure of the β-phase is not of maximal density, it may be assumed that the found hardness value is not a limit value for boron (III) oxide. According to the approach proposed in [7], the highest hardness (30 GPa) is expected for the γ-$B_2O_3$ hypothetic isotropic dense phase with the structure of corundum (see Fig. 1c).

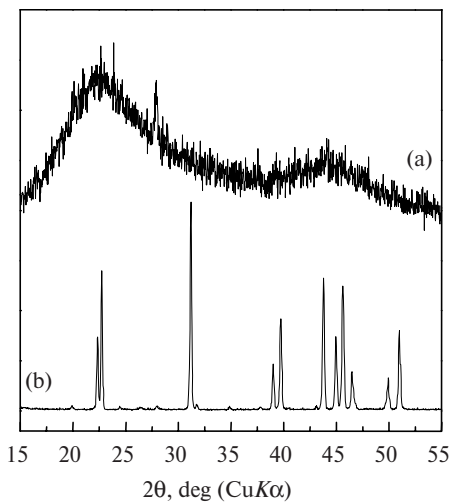

**Fig. 2**. Diffraction patterns of glass-like $B_2O_3$ (a) and β-$B_2O_3$ (b).

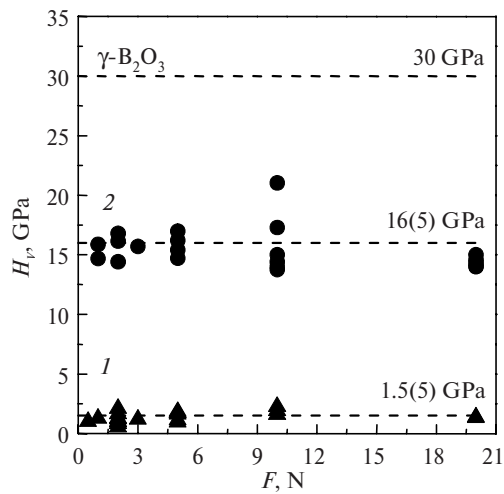

**Fig. 3**. Hardness of glass-like $B_2O_3$ (*1*) and β-$B_2O_3$ (*2*) vs. loading.

We would like to thank Agence Nationale de la Recherche (France) for the financial support (grant NT05-3_42601).